\documentclass[twocolumn, preprintnumbers,prl,aps,amssymb,showpacs,superscriptaddress]{revtex4}
\usepackage{graphicx}
\usepackage{natbib}
\usepackage{dcolumn}
\usepackage{wasysym}
\usepackage{bm} \usepackage{color}
\begin{document}

\newcommand{\ie}{{\it i.e.}}
\newcommand{\eg}{{\it e.g.}}
\newcommand{\etal}{{\it et al.}}

\newcommand{\K}{Ba$_{1-x}$K$_x$Fe$_2$As$_2$}
\newcommand{\KFeAs}{KFe$_2$As$_2$}
\newcommand{\Co}{Ba(Fe$_{1-x}$Co$_x$)$_2$As$_2$}
\newcommand{\Kzero}{$\kappa_0/T$}
\newcommand{\KN}{$\kappa_{\rm N}/T$}

\newcommand{\Tc}{$T_{\rm c}$}
\newcommand{\Hc}{$H_{\rm c2}$}
\newcommand{\Hstar}{$H^\star$}
\newcommand{\vF}{$v_{\rm F}$}

\newcommand{\units}{$\mu \text{W}/\text{K}^2\text{cm}$}
\newcommand{\p}[1]{\left( #1 \right)}
\newcommand{\Dd}[2]{\frac{\text{d} #1}{\text{d}#2}}


\title{
Thermal Conductivity of the Iron-Based Superconductor FeSe :\\
 Nodeless Gap with Strong Two-Band Character
}


\author{P.~Bourgeois-Hope} 
\affiliation{D\'epartement de physique \& RQMP, Universit\'e de Sherbrooke, Sherbrooke, Qu\'ebec J1K 2R1, Canada}

\author{S.~Chi} 
\affiliation{Department of Physics \& Astronomy, University of British Columbia, Vancouver, British Columbia V6T 1Z1, Canada}

\author{D.~A.~Bonn} 
\affiliation{Department of Physics \& Astronomy, University of British Columbia, Vancouver, British Columbia V6T 1Z1, Canada}
\affiliation{Canadian Institute for Advanced Research, Toronto, Ontario M5G 1Z8, Canada}

\author{R.~Liang} 
\affiliation{Department of Physics \& Astronomy, University of British Columbia, Vancouver, British Columbia V6T 1Z1, Canada}
\affiliation{Canadian Institute for Advanced Research, Toronto, Ontario M5G 1Z8, Canada}

\author{W.~N.~Hardy} 
\affiliation{Department of Physics \& Astronomy, University of British Columbia, Vancouver, British Columbia V6T 1Z1, Canada}
\affiliation{Canadian Institute for Advanced Research, Toronto, Ontario M5G 1Z8, Canada}

\author{T.~Wolf} 
\affiliation{Institute of Solid State Physics (IFP), Karlsruhe Institute of Technology, D-76021, Karlsruhe, Germany}

\author{C.~Meingast} 
\affiliation{Institute of Solid State Physics (IFP), Karlsruhe Institute of Technology, D-76021, Karlsruhe, Germany}

\author{N.~Doiron-Leyraud} 
\affiliation{D\'epartement de physique \& RQMP, Universit\'e de Sherbrooke, Sherbrooke, Qu\'ebec J1K 2R1, Canada}

\author{Louis Taillefer}
\altaffiliation{E-mail: louis.taillefer@usherbrooke.ca }
\affiliation{D\'epartement de physique \& RQMP, Universit\'e de Sherbrooke, Sherbrooke, Qu\'ebec J1K 2R1, Canada}
\affiliation{Canadian Institute for Advanced Research, Toronto, Ontario M5G 1Z8, Canada}

\date{\today}


\begin{abstract}

The thermal conductivity $\kappa$ of the iron-based superconductor FeSe was measured at temperatures down to 50~mK in magnetic fields up to 17~T. 
In zero magnetic field, the electronic residual linear term in the $T=0$ limit, $\kappa_0/T$, is vanishingly small. 
Application of a magnetic field $H$ causes no increase in $\kappa_0/T$ initially. 
Those two observations show that there are no zero-energy quasiparticles that carry heat and therefore no nodes in the superconducting gap of FeSe. 
The full field dependence of $\kappa_0/T$ has the classic shape of a two-band superconductor, such as MgB$_2$:
it rises exponentially at very low field, with a characteristic field $H^{\star} <<$~\Hc, and then more slowly up to the upper critical field \Hc.
This shows that the superconducting gap is very small on one of the pockets in the Fermi surface of FeSe. 


\end{abstract}


\maketitle


In the family of iron-based superconductors, the simple binary material FeSe
has attracted much attention because when made in thin-film form 
its superconductivity appears to persist up to critical temperatures as high as \Tc~$\simeq 100$~K \cite{Tc100K}.
In bulk form, FeSe is unusual in
that it undergoes the standard tetragonal-to-orthorhombic structural transition without the
usual accompanying antiferromagnetic transition  \cite{McQueen2009,Bendele2010}.
This raises fundamental questions about the role of magnetism in causing
superconductivity and nematicity.

A basic property of any superconductor is its gap function or structure,
which is related to the symmetry of its pairing state,
yet there is no consensus on the gap structure of FeSe.
A thermal conductivity study of non-stoichiometric FeSe$_x$ 
revealed no nodes in the gap \cite{DongPRB2009}.
By contrast, a huge residual linear term at $T \to 0$
was reported in a subsequent study of thermal conductivity in stoichiometric FeSe \cite{KasaharaPNAS2014},
viewed as evidence of nodes in the gap.
STM measurements detect a V-shaped density of states at low energy \cite{KasaharaPNAS2014,Watashige2015} and
the penetration depth has a nearly linear temperature dependence at low temperature \cite{KasaharaPNAS2014},
both features interpreted in terms of nodes.
%
%
Specific heat measurements down to $T = 0.5$~K show that there are low-lying excitations, 
but the data cannot distinguish clearly between nodes or just a small minimum gap \cite{Lin2011}.
Calculations for a model where pairing proceeds via spin excitations yield
a superconducting gap with accidental nodes on one of the Fermi surface pockets \cite{Kreisel2015}.

In this Letter, we investigate the gap structure of pure stoichiometric FeSe using 
thermal conductivity, 
a bulk probe of the superconducting gap highly sensitive to the presence
or absence of nodes~\cite{Shakeripour2009}. 
Measurements were performed on two single crystals,
grown by two different groups, and the results are in excellent agreement.
We find that the residual linear term in $\kappa(T)$ as $T \to 0$, \Kzero, is negligible
at $H = 0$ and it rises only slowly at first with magnetic field $H$,
clear evidence that there are no nodes in the gap, in contrast
with the prior study \cite{KasaharaPNAS2014}.
However, the field dependence reveals the presence of a very small gap
on some part of the Fermi surface, which could account for the 
low-energy quasiparticle excitations detected in STM and penetration depth.


\begin{figure}[t]
\centering
\includegraphics[width=8.8cm]{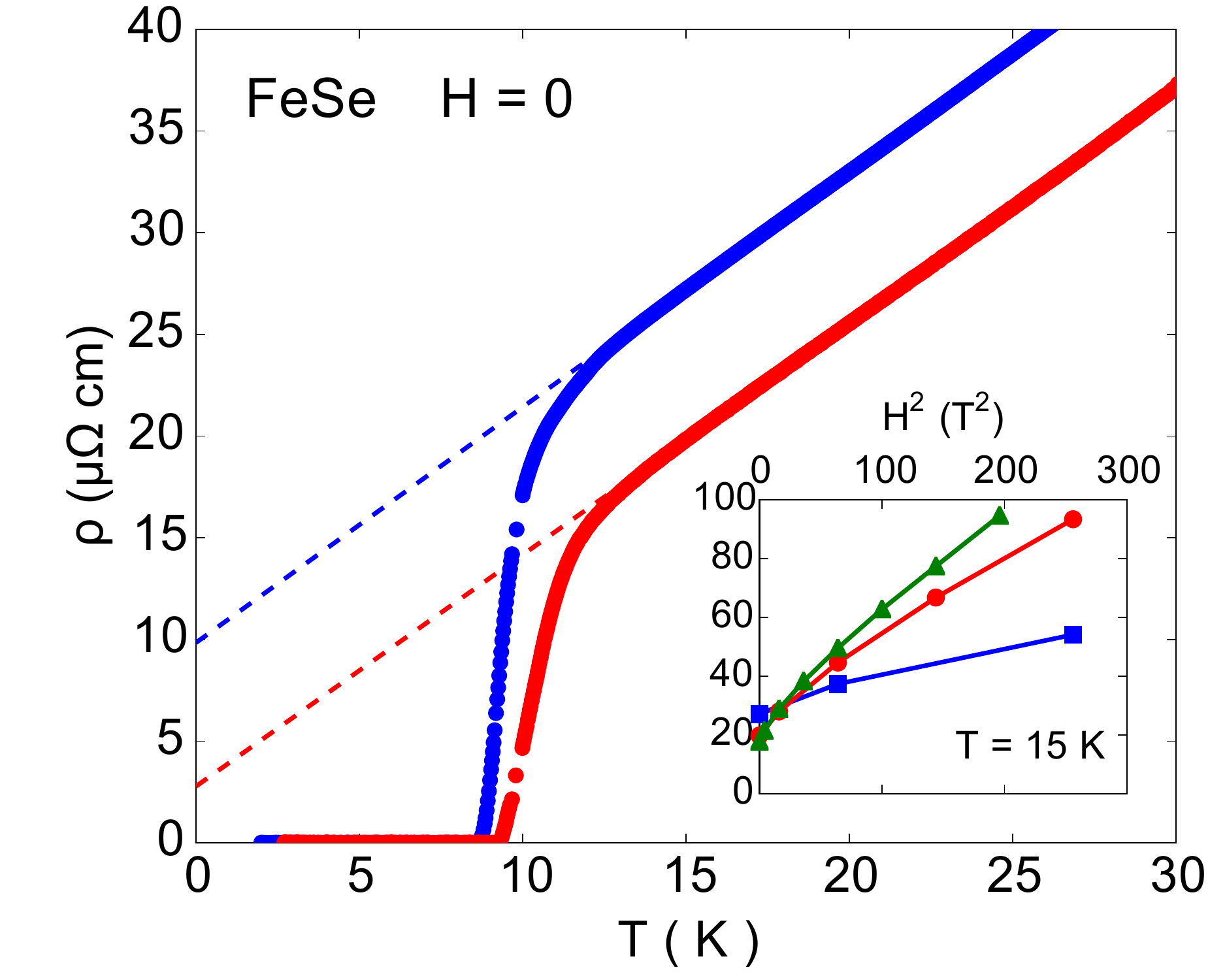}
\caption{
In-plane electrical resistivity $\rho(T)$ of FeSe for our samples A (red) and B (blue).
The dashed lines are a linear fit to $\rho(T)$ between 15~K and 20~K, extended to $T=0$,
giving a residual resistivity $\rho(T \to 0) = 2.8~\mu \Omega$~cm (A) and 
$9.8~\mu \Omega$~cm (B).
{\it Inset}:
Dependence of $\rho$~on magnetic field $H$, at $T = 15$~K, plotted as $\rho$~vs $H^2$,
for samples A (red circles) and B (blue squares), 
compared with corresponding data in ref.~\onlinecite{KasaharaPNAS2014}~(green triangles).
}
\label{Fig1}
\end{figure}



\begin{figure*}[t]
\centering
\includegraphics[width=17.5cm]{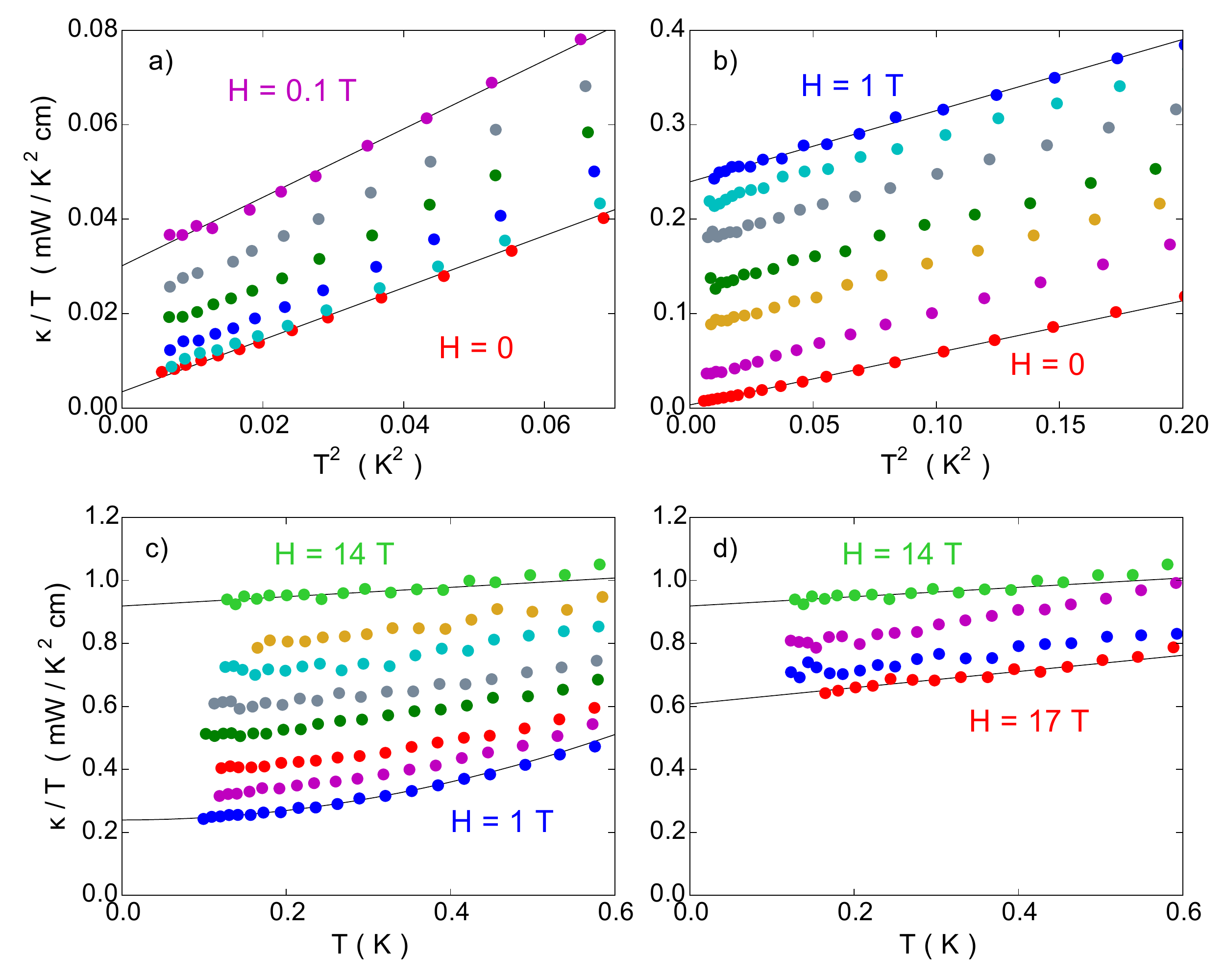}
\caption{
Temperature dependence of the in-plane thermal conductivity $\kappa(T)$ of FeSe, measured on sample A, 
with an applied magnetic field $H \parallel c$.
(a)
Plotted as $\kappa/T$ vs $T^2$ for $H = 0$, 0.02, 0.04, 0.06, 0.08, and 0.1~T (from bottom to top).
Lines are a fit to $\kappa/T = a + bT^2$, used to obtain the residual linear term at $T=0$, $a \equiv \kappa_0/T$. 
(b)
Plotted as $\kappa/T$ vs $T^2$ for $H = 0$, 0.1, 0.2, 0.3, 0.5, 0.75, and 1.0~T (from bottom to top).
Lines are a fit to $\kappa/T = a + bT^2$.
(c)
Plotted as $\kappa/T$ vs $T$ for $H = 1$, 2, 4, 8, 10, 12, 13, and 14~T (from bottom to top).
The lower line is a fit to $\kappa/T = a + bT^2$ ($H = 1$~T) and the upper line a fit to $\kappa/T = a + bT$ ($H = 14$~T).
(d)
Plotted as $\kappa/T$ vs $T$ for $H = 14$, 15, 16, and 17~T (from top to bottom).
Lines are a fit to $\kappa/T = a + bT$. 
The values of $a \equiv$~\Kzero~obtained from a fit to either $\kappa/T = a + bT^2$ ($H < 11$~T) or $\kappa/T = a + bT$ ($H > 11$~T)  
are plotted as \Kzero~vs $H$~in Fig.~3.
}
\label{Fig2}
\end{figure*}



{\it Methods.--}
Single crystals of FeSe were grown by vapour transport \cite{FeSe-Growth}.
Sample A was grown at UBC in Vancouver;
sample B was grown at KIT in Karlsruhe.
They have similar characteristics, but sample A is slightly cleaner,
resulting in a slightly higher \Tc, namely 9.3~K (A) vs 8.6~K (B)~(Fig.~1).
The contacts to the sample were made using silver epoxy.
The thermal conductivity was measured in a dilution refrigerator
down to 50~mK, for a heat current in the basal plane
of the orthorhombic crystal structure, as described elsewhere \cite{Reid2010}.
A magnetic field up to 17~T was applied along the $c$ axis,
and always changed at $T >$~\Tc.


{\it Resistivity.--}
The in-plane resistivity $\rho(T)$ of our two samples is in excellent agreement with previously published data \cite{KasaharaPNAS2014},
when normalized to a common value at $T = 300$~K, namely $\rho(300~{\rm K}) = 410~\mu \Omega$~cm.
Differences are only visible when zooming at low temperature, as done in Fig.~1.
We see that the curve for sample B is shifted up relative to that of sample A,
so that 
$\rho(T = 15~{\rm K}) = 20.0~\mu \Omega$~cm (A) and $25.5~\mu \Omega$~cm (B).
For comparison, the sample of FeSe studied in ref.~\onlinecite{KasaharaPNAS2014} 
has \Tc~$\simeq 9.4$~K and $\rho(T = 15~{\rm K}) \simeq 18~\mu \Omega$~cm,
 showing that its disorder level is similar to, but slightly lower than that of sample A.
 A linear extrapolation of $\rho(T)$ to $T=0$ yields
$\rho(T \to 0) = 2.8~\mu \Omega$~cm (A) and $9.8~\mu \Omega$~cm (B).
This gives a residual resistance ratio RRR~$\equiv \rho(300~{\rm K}) / \rho(T \to 0) = 148$~(A) and 42 (B),
a simple dimensionless measure of sample quality.
By comparison, the sample of FeSe$_x$ in ref.~\onlinecite{DongPRB2009} has RRR~$< 10$.

Owing to its semi-metal-like Fermi surface made of small hole-like and electron-like pockets,
FeSe displays a strong orbital magnetoresistance (MR)~\cite{Watson2015},
which goes approximately as $\rho(T \to 0) \propto H^2$.
The level of disorder (impurities) is likely to affect the magnitude of the MR,
and the cleaner the sample the larger the MR.
In the inset of Fig.~1, we compare the MR measured just above \Tc~(at $T = 15$~K)
in our two samples, and that of ref.~\onlinecite{KasaharaPNAS2014}.
As expected, the MR increases with increasing RRR, and 
we see that the sample of ref.~\onlinecite{KasaharaPNAS2014} is slightly cleaner than
our sample~A.


\begin{figure}[t]
\centering
\includegraphics[width=8.5cm]{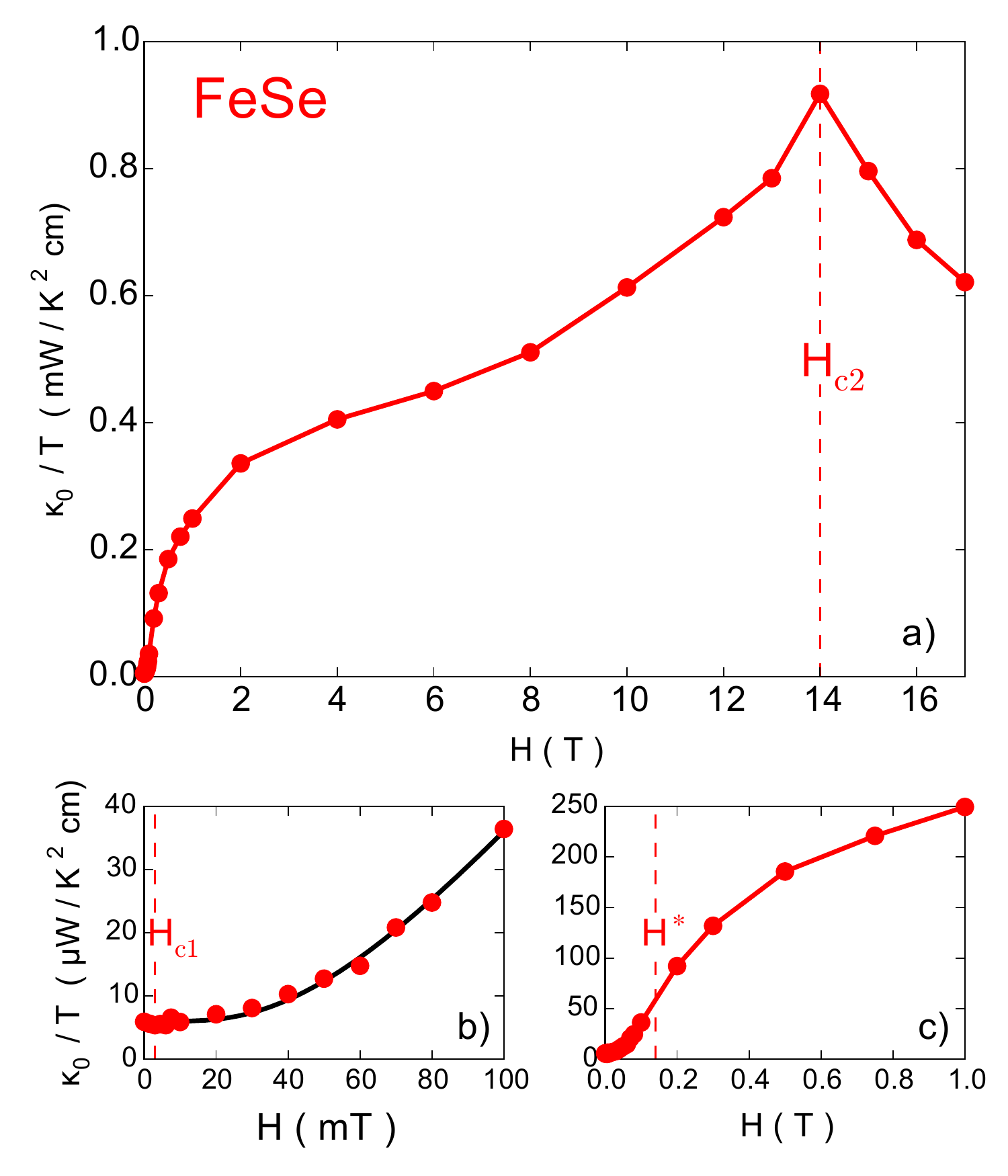}
\caption{
Field dependence of the residual linear term \Kzero~in FeSe,
obtained from fits to the data as in Fig.~2. 
a)
Over the full field range.
The vertical dashed line marks the upper critical field, \Hc~$=14$~T.
b)
Zoom below $H = 0.1$~T.
The vertical dashed line marks the lower critical field, $H_{\rm c1} = 3$~mT~\cite{Abdel-Hafiez2013}.
The full line is an exponential fit to the data up to 0.1~T.
c)
Zoom below $H = 1.0$~T.
The vertical dashed line marks the inflexion point from upward to downward curvature, at \Hstar~$\simeq 150$~mT.
}
\label{Fig3}
\end{figure}



{\it Thermal conductivity.--}
The thermal conductivity $\kappa(T)$ of FeSe at low temperature 
is shown in the four panels of Fig.~2, for 26 different values of the magnetic field $H$,
ranging from $H = 0$ to $H = 17$~T. 
Data taken at $H = 1.5$, 3.0, 4.5, 6.0, 7.5~mT are not shown,
as they are essentially identical to the data at $H = 0$ and $H = 0.01$~T.
At low field ($H < 1.0$~T), the data are well described by the form
$\kappa / T = a + bT^2$ below $T \simeq 0.4$~K (Figs.~2a, 2b).
The residual linear term, $a \equiv \kappa_0/T$, is purely electronic, 
and the second term, $bT^2$, is due to phonon conduction \cite{Shakeripour2009}.
In that regime, phonons are mostly scattered by the sample boundaries, and the phonon mean free path is constant.
In this Letter, our focus is entirely on $\kappa_0/T$, the electronic transport due to zero-energy 
quasiparticles. 
At higher field, $\kappa(T)$ gradually becomes more linear (Fig.~2c), as in the normal state
above \Hc~$= 14$~T (Fig.~2d).
In that regime, phonons are predominantly scattered by electrons, and their mean free path goes as $1/T$.
Above 10~T, a fit to the form $\kappa / T = a + bT$ below $T \simeq 0.4$~K is used to extract
\Kzero~(Figs.~2c, 2d).

At $H = 0$, a fit of the data below 0.4~K to the form $\kappa / T = a + bT^2$ yields a very small
value of $a$, of magnitude $6 \pm 2~\mu$W / K$^2$~cm (Fig.~2a).
To put it in perspective, this value should be compared to the value in the normal state, \KN, which we estimate by applying the Wiedemann-Franz law
to the residual resistivity $\rho(T \to 0) = 2.8~\mu \Omega$~cm (Fig.~1), 
giving 
\KN~$= (\pi^2/3) (k_{\rm B}/e)^2 /  \rho(T \to 0) = 8.8$~mW / K$^2$~cm.
We see that $a = \kappa_0/T$ is less than 0.1~\%~of the normal state conductivity,
a negligible value.
This shows that there are no zero-energy quasiparticles in the superconducting state of FeSe
at $H=0$.

Applying a magnetic field is a controlled way of exciting quasiparticles in the superconducting ground state at $T=0$.
Looking at the full $H$ dependence of \Kzero~up to 17~T~(Fig.~3a),
we see the typical behaviour of a two-band superconductor like MgB$_2$ \cite{Sologubenko2002}
or NbSe$_2$ \cite{Boaknin2003}.
Two features are striking.
The first is the sharp cusp at $H = 14$~T.
This is the upper critical field \Hc, below which vortices appear in the sample.
The appearance of vortices introduces an additional scattering process,
which suddenly curtails the mean free path, causing an abrupt drop in conductivity below \Hc,
in samples with long electronic mean free path.
This happens in any clean type-II superconductor, whether the gap is nodeless -- as in~Nb
or LiFeAs \cite{Tanatar2011} -- or nodal -- as in KFe$_2$As$_2$ \cite{Reid2012}
or YBa$_2$Cu$_3$O$_y$ \cite{Grissonnanche2014},
provided the elastic normal-state mean free path is much longer than the $T=0$ coherence 
length $\xi$ (\ie~the inter-vortex separation at \Hc).

Note that the decrease in \Kzero~above \Hc~(Fig.~3a) is due to the strong magnetoresistance of
the normal state (inset of Fig.~1).
As discussed below, the $H$ dependence of \Kzero~is in quantitative agreement with the known 
$H$ dependence of $\rho$ in the normal state \cite{Watson2015}. 
This proves that the cusp indeed corresponds to the end of the
vortex state and it rules out its previous interpretation as an internal phase transition inside the vortex state \cite{KasaharaPNAS2014}.


\begin{figure}[t]
\centering
\includegraphics[width=8.5cm]{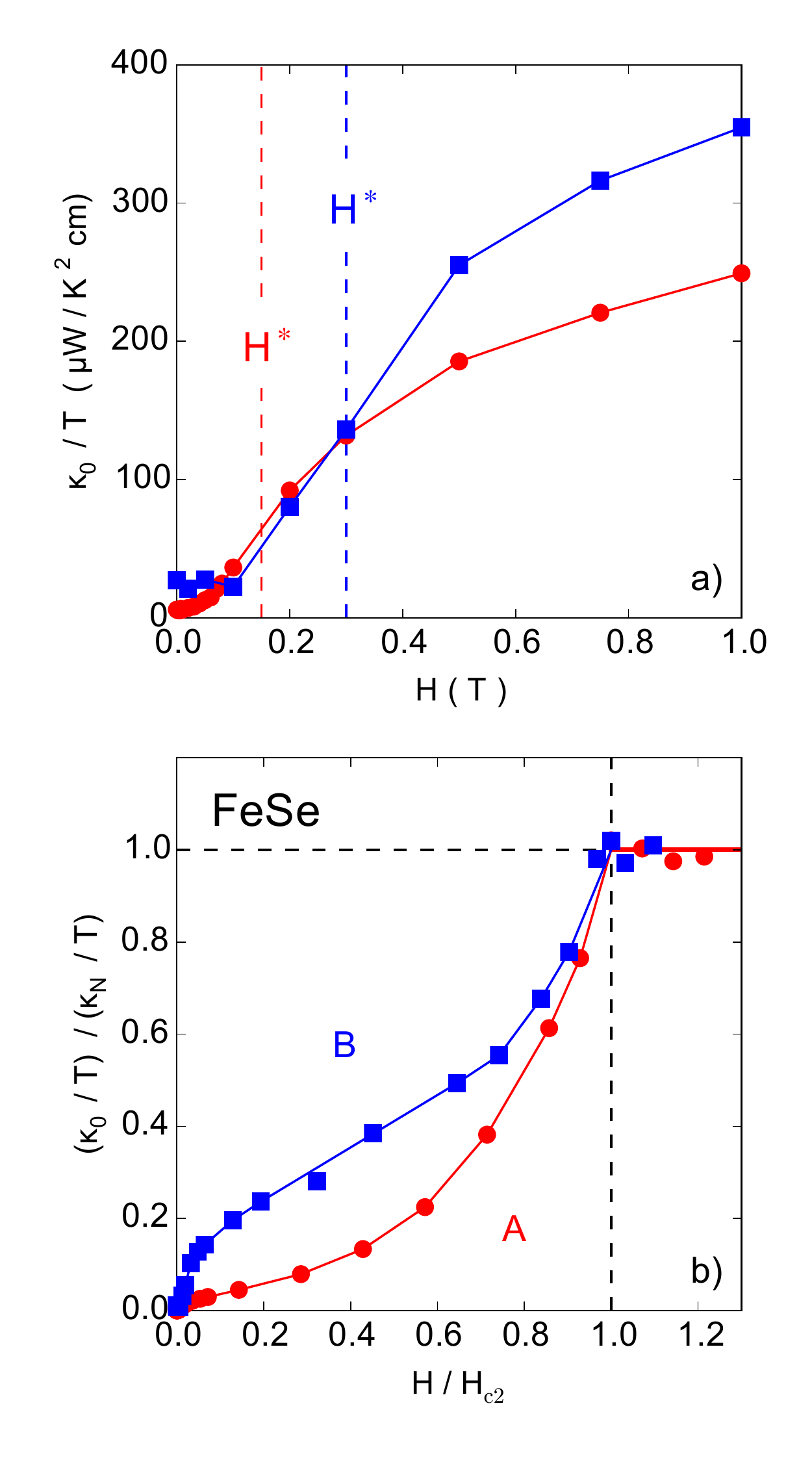}
\caption{
Field dependence of the residual linear term \Kzero~in FeSe,
comparing sample A (red circles) and sample B (blue squares). 
a)
Zoom of the raw data below $H = 1.0$~T.
The color-coded vertical dashed lines mark the approximate location of the inflexion point, 
separating a regime of upward curvature at low $H$ and a regime of downward curvature at higher $H$,
at \Hstar~$\simeq 0.15$~T (red) and \Hstar~$\simeq 0.3$~T (blue).
b)~Over the full field range, with \Kzero~normalized by the field-dependent normal-state conductivity \KN~(see text)
and $H$ normalized by \Hc.
}
\label{Fig4}
\end{figure}


The second striking feature of \Kzero~vs $H$~is the rapid rise at low $H$~(Fig.~3a).
To investigate this closely, the field was increased in very small steps,
starting with $H=$~1.5~mT, then 3.0~mT, and so on (Fig.~3b).
In FeSe, the lower critical field above which vortices first enter the sample at $T \to 0$
is $H_{\rm c1} \simeq 3$~mT \cite{Abdel-Hafiez2013}.
We find that increasing $H$ up to 20~mT, a field 7 times larger than $H_{\rm c1}$, 
causes no increase in quasiparticle conduction. 
This confirms that there are no nodes in the gap, for if there were, a field greater than $H_{\rm c1}$
would rapidly excite nodal quasiparticles.

In Fig.~3b, 
we see that the initial rise in \Kzero~vs $H$ is exponential,
so that the field-induced quasiparticle heat conduction in FeSe is an activated process
(at low $H$),
very different from the rapid rise characteristic of nodal superconductors \cite{Shakeripour2009}.
This shows that there is a minimum gap for quasiparticle excitations,
\ie~there are no nodes anywhere in the gap structure.
However, that minimum gap is much smaller than the maximum gap responsible
for setting \Hc.
Indeed, the characteristic field for the initial rise is roughly \Hstar~$\simeq$~\Hc/100,
if we define \Hstar~as the inflexion point in \Kzero~vs $H$, where \Kzero~goes from a positive 
to a negative curvature (Fig.~3c),
giving \Hstar~$\simeq 0.15$~T.

The quantity that controls how fast \Kzero~rises with $H$ is not the superconducting gap $\Delta$~but
the coherence length $\xi \propto v_{\rm F} / \Delta$, where \vF~is the Fermi velocity.
In a single-band situation, the upper critical field is set by $\xi$: 
\Hc~$\propto 1 / \xi^2 \propto (\Delta / v_{\rm F})^2$.
In a two-band model, the Fermi surface with the smaller $\xi$ will set \Hc,
while the surface with the larger $\xi$ will control \Hstar.
In the two-band superconductor MgB$_2$, \Hstar~$\simeq$~\Hc/10 because the small gap is 
3 times smaller than the large gap \cite{Sologubenko2002}.

If we accept the interpretation that the Fermi surface of FeSe consists of two distinct pockets -- a small $\Gamma$-centred 
hole pocket and a small electron pocket at the corner of the Brillouin zone \cite{KasaharaPNAS2014,Terashima2014} --
with comparable values of $v_{\rm F}$ \cite{Terashima2014},
then we are forced to conclude that the superconducting gap varies by an order of magnitude 
around the Fermi surface.
There are two scenarios for the $k$ dependence of the gap in FeSe.
In the first, the gap is roughly isotropic on both surfaces but much smaller on one of them,
\ie~$\Delta^\star << \Delta$.
In the second, the gap is highly anisotropic around one of the surfaces, perhaps both,
varying from a minimum value $\Delta_{\rm min}$ to a maximum value $\Delta_{\rm max}$,
such that $\Delta_{\rm min} << \Delta_{\rm max}$.
Note that in the iron arsenide \K~the superconducting gap acquires a deeper and deeper 
minimum as the material is more and more underdoped, with $x$ decreasing below $x = 0.4$
(where \Tc~is maximal)~\cite{Reid2016}.
Of course, it may be that both effects are present in pure FeSe: 
the gap is smaller {\it and} anisotropic on one of the two Fermi surfaces.
Note that because FeSe is orthorhombic the superconducting gap function 
will involve a mixture of $s$-wave and $d$-wave components, which 
naturally introduces anisotropy~\cite{Kang2014}.

{\it Effect of disorder.--}
It is instructive to compare samples with  different levels of disorder.
In Fig.~4a, we plot \Kzero~vs $H$, for samples A and B, at fields below $H = 1.0$~T.
We see in sample B the same characteristics we saw in sample A, 
namely a negligible \Kzero~at $H = 0$ ($\simeq 1$\% of \KN), 
a flat \Kzero~at low $H$
(in this case up to $H \simeq 100$~mT), 
and a strong two-band character,
with an inflexion point now at \Hstar~$\simeq 0.3$~T.
So sample B leads us to the same qualitative conclusions:
no nodes, but a very small gap on part of the Fermi surface.
Quantitatively, the minimum gap appears to be larger, as measured by 
the larger value of \Hstar. 
It therefore seems that disorder enhances the minimum gap.
One way to interpret this effect is for disorder to reduce (or remove) the anisotropy present in the small gap.
It is then conceivable that in samples cleaner than sample A, the anisotropy is such that the deep minima further deepen to
produce shallow accidental nodes where the gap changes sign~\cite{Mishra2009}.
This is the mechanism proposed by Kasahara and co-workers~\cite{KasaharaPNAS2014} to 
explain why they see a large residual \Kzero~in their clean sample of FeSe when
no such residual term had been detected in the more disordered FeSe$_x$~\cite{DongPRB2009}.
Note, however, that the value of \Kzero~they report at $H = 0$ is enormous \cite{KasaharaPNAS2014},
20-40 times larger than the value \Kzero~$\simeq 100$-200~mW / K$^2$ cm we observe in both our samples
just above \Hstar~(Fig.~4a), and a sizeable fraction of the total normal-state conductivity \KN.
Their enormous \Kzero~value at $H=0$ remains a puzzle.

In Fig.~4b, \Kzero~is normalized by the field-dependent normal-state conductivity 
\KN, 
and $H$ is normalized by the zero-temperature value of \Hc~(essentially the same in the two samples).
We obtain \KN~by fitting the data points above \Hc~(the 4 data points at $H \geq 14$~T in Fig.~3a, for sample A) 
to the relation \KN~$= L_0 / (a + b H^2)$, 
since $\rho = a + b H^2$ in FeSe \cite{Watson2015}, given that  
the Wiedemann-Franz law requires that 
\KN~$= L_0 /  \rho$ in the $T = 0$ limit, with $L_0 \equiv (\pi^2/3) (k_{\rm B}/e)^2$.
For sample A, we obtain $b \simeq 50$~n$\Omega$~cm / T$^2$, in excellent agreement
with the data of ref.~\onlinecite{Watson2015} for the normal-state MR of FeSe.
In Fig.~4b, we observe a clear shoulder in (\Kzero) / (\KN)~at $H/$\Hc~$\simeq 1 / 20$,
very similar to the shoulder seen in MgB$_2$ at $H/$\Hc~$\simeq 1 / 9$.
In this normalized plot, the two-band character is more apparent 
for sample B than for sample A.
The disorder in sample B is such as to degrade more effectively than in sample A
the conductivity of the large-gap Fermi surface relative to that of the small-gap Fermi surface.
Our normalized data for \Kzero~vs $H$ on sample B and A (Fig.~4b) can be viewed as cleaner and much cleaner versions
of the data in FeSe$_x$ \cite{DongPRB2009}, respectively, but they bear no resemblance to 
the data of ref.~\cite{KasaharaPNAS2014}.


{\it Summary.--}
In summary, our thermal conductivity measurements on two high-quality crystals of FeSe
reveal a superconducting gap without nodes, but with a strong two-band character, 
whereby the gap magnitude on one pocket of the Fermi surface of FeSe is 
an order of magnitude smaller than its magnitude on the other pocket. 
The presence of such a small gap will make various superconducting properties
of FeSe, such as the penetration depth, appear as though they come from a nodal gap, 
unless measurements are carried out  to very low temperature and/or very low energy.


{\it Acknowledgements.--}
The work at Sherbrooke was supported by a 
Canada Research Chair,
the Canadian Institute for Advanced Research, 
the National Science and Engineering Research Council of Canada, 
the Fonds de recherche du Qu\'ebec - Nature et Technologies, 
and the Canada Foundation for Innovation.
%
%



\begin{references}

\bibitem{Tc100K}
J.-F.~Ge {\it et al.},
Nature Materials {\bf 14}, 285 (2015).

\bibitem{McQueen2009}
M.~McQueen {\it et al.}, 
Phys. Rev. Lett. {\bf 103}, 057002 (2009).
 
\bibitem{Bendele2010}
 M.~Bendele {\it et al.}, 
Phys. Rev. Lett. {\bf 104}, 087003 (2010).
 
\bibitem{DongPRB2009}
J.~K.~Dong {\it et al.},
Phys. Rev. B {\bf 80}, 024518 (2009).

\bibitem{KasaharaPNAS2014}
S.~Kasahara {\it et al.},
PNAS {\bf 111}, 16309 (2014).
 
\bibitem{Watashige2015}
 T.~Watashige {\it et al.}, 
Phys. Rev. X {\bf 5}, 031022 (2015).
 
\bibitem{Lin2011}
J.-Y.~Lin {\it et al.},
Phys. Rev. B {\bf 84}, 220507(R) (2011).

\bibitem{Kreisel2015}
A.~Kreisel {\it et al.},
Phys. Rev. B {\bf 92}, 224515 (2015).

\bibitem{Shakeripour2009} 
H. Shakeripour {\it et al.},
New J. Phys. {\bf 11}, 055065 (2009).

\bibitem{FeSe-Growth}
A.~E.~Boehmer  {\it et al.},
Phys. Rev. B {\bf 87}, 180505(R) (2013).

\bibitem{Reid2010}
J.~Ph. Reid {\it et al.},
Phys. Rev. B {\bf 82}, 064501 (2010).

\bibitem{Watson2015} 
W.D.~Watson {\it et al.}, 
Phys. Rev. Lett. {\bf 115}, 027006 (2015).

\bibitem{Abdel-Hafiez2013}
M.~Abdel-Hafiez {\it et al.}, 
Phys. Rev. B { \bf 88}, 174512 (2013).

\bibitem{Sologubenko2002}
A.V.~Sologubenko {\it et al.}, 
Phys. Rev. B { \bf 66}, 014504 (2002).

\bibitem{Boaknin2003} 
E.~Boaknin {\it et al.}, 
Phys. Rev. Lett. {\bf 90}, 117003 (2003).

\bibitem{Tanatar2011} 
M.~A.~Tanatar {\it et al.},
Phys. Rev. B {\bf 84}, 054507 (2011).

\bibitem{Reid2012} 
J.~Ph.~Reid {\it et al.},
Phys. Rev. Lett. {\bf 109}, 087001 (2012).

\bibitem{Grissonnanche2014} 
G.~Grissonnanche {\it et al.},
Nat. Commun. {\bf 5}, 3280 (2014).

\bibitem{Terashima2014} 
T.~Terashima {\it et al.},
Phys. Rev. B {\bf 90}, 144517 (2014).

\bibitem{Reid2016} 
J.~Ph.~Reid {\it et al.},
arXiv:1602.03914 (2016).

\bibitem{Kang2014} 
J.~Kang {\it et al.},
Phys. Rev. Lett. {\bf 113}, 217001 (2014).

\bibitem{Mishra2009} 
V.~Mishra {\it et al.},
Phys. Rev. B {\bf 79}, 094512 (2009).



\end{references}
\end{document}